\documentclass[twocolumn, reprint, amsmath,amssymb,aps,prb]{revtex4-2}

\usepackage{graphicx}
\usepackage{dcolumn}
\usepackage{bm,epsfig}
\usepackage{epstopdf}
\usepackage{xcolor}

\begin{document}

\title{Ground state of Ce$_{3}$Bi$_{4}$Pd$_{3}$ unraveled by hydrostatic pressure}

\author{M. O. Ajeesh$^{1}$, S. M. Thomas$^{1}$, S. K. Kushwaha$^{2}$, E. D. Bauer$^{1}$, F. Ronning$^{1}$, J. D. Thompson$^{1}$, N. Harrison$^{2}$, and P. F. S. Rosa$^{1}$}
\affiliation{
$^{1}$  Los Alamos National Laboratory, Los Alamos, New Mexico 87545, U.S.A.\\
$^{2}$ National High Magnetic Field Laboratory, Los Alamos, New Mexico 87545, U.S.A.}
\date{\today}

\begin{abstract}

Noncentrosymmetric Ce$_{3}$Bi$_{4}$Pd$_{3}$ has attracted a lot of attention as a candidate for strongly correlated topological material, yet its experimental ground state remains a matter of contention. Two conflicting scenarios have emerged from a comparison to prototypical Kondo insulator Ce$_{3}$Bi$_{4}$Pt$_{3}$: either Ce$_{3}$Bi$_{4}$Pd$_{3}$ is a spin-orbit-driven topological semimetal or a Kondo insulator with smaller Kondo coupling than its Pt counterpart. Here we determine the ground state of Ce$_{3}$Bi$_{4}$Pd$_{3}$ via electrical resistivity measurements under hydrostatic pressure, which is a clean symmetry-preserving tuning parameter that increases hybridization but virtually preserves spin-orbit coupling. Ce$_{3}$Bi$_{4}$Pd$_{3}$ becomes more insulating under pressure, which is a signature of Ce-based Kondo insulating materials. Its small zero-pressure gap increases quadratically with pressure, 
similar to the behavior observed in the series Ce$_{3}$Bi$_{4}$(Pt$_{1-x}$Pd$_{x}$)$_{3}$, which indicates that Pt substitution and applied pressure have a similar effect. Our result not only demonstrates that Kondo coupling, rather than spin-orbit coupling, is the main tuning parameter in this class of materials, but it also establishes that Ce$_{3}$Bi$_{4}$Pd$_{3}$ has a narrow-gap Kondo insulating ground state.

 \end{abstract}

\maketitle

 Topological classification schemes
have uncovered unprecedented quantum states of matter by employing a bulk invariant to distinguish states that share the same symmetries~\cite{Kane05a,Kane05b,Moore07,Roy09}. 
A prototypical example of a topological state of matter is the 
fractional quantum Hall state, which exhibits a nonlocal order parameter, fractionalized excitations, and long-range entanglement~\cite{Konig08,Hasan10}.  
For three-dimensional insulators, the simplest case of inversion-symmetric bulk materials utilizes the so-called $Z_{2}$ topological invariant,
which is defined as the product of the parities of the occupied bands at high-symmetry points in momentum space, to determine 
whether the material is topologically nontrivial~\cite{Fu07a,Fu07b,Dzero16}. A topological insulating state stems from the combination of 
  spin-orbit interactions and time-reversal symmetry, and its edge state is typically a manifestation of gapless excitations at 
  the interface between topologically distinct regions~\cite{Hasan10, Qi11}.

The concept of band topology can also be extended beyond insulating and inversion-symmetric states, which has enabled the discovery of new classes of topological matter~\cite{Young12,Wang12,Burkov16,Lv21,Arm18}.
In particular, a Weyl Kondo semimetal (WKSM) phase has recently been predicted to emerge
in strongly correlated materials with broken inversion symmetry~\cite{Lai18, Chang18}. In a WKSM, the Kondo effect gives rise to
 strongly renormalized Weyl nodes in the bulk and Fermi arcs on the surface. Notably, the large renormalization factor
 of nodes pinned to the Fermi energy 
 is predicted to enable thermodynamic quantities, such as specific heat, to experimentally probe Weyl nodes in these materials.

Noncentrosymmetric Ce$_{3}$Bi$_{4}$Pd$_{3}$ has been put forward as a prime Weyl Kondo semimetal candidate~\cite{Lai18, Cao20}, yet its ground state remains a matter of contention.
On one hand, the electronic contribution to the specific heat includes a linear-in-temperature term as well as a $\Gamma T^{3}$ term whose prefactror $\Gamma$ is consistent with the theoretical WKSM proposal~\cite{Dzsaber17, Lai18}. In addition, the weak temperature dependence of electrical resistance reported initially was taken as an indication of the absence of a well-defined energy gap~\cite{Dzsaber17}. Furthermore, recent experiments at zero magnetic 
field reveal a giant spontaneous Hall effect, which is explained as a substantial Berry curvature contribution coming from 
tilted renormalized Weyl nodes~\cite{Dzsaber21}. On the other hand, independent experimental reports argued for a small-gap 
insulating ground state based on activated behavior in electrical resistivity and Hall resistivity as well as 
fits of the low-temperature specific heat to a Schotte-Schotte anomaly~\cite{Kushwaha19}. In addition, a broad maximum in magnetic susceptibility at
about $T_{M} = 5$~K was attributed to a Kondo gap scale of 15-20~K that can be suppressed with applied magnetic fields. 

To shed light on this issue, a comparison to prototypical Kondo insulator Ce$_{3}$Bi$_{4}$Pt$_{3}$~\cite{Hundley90, Hundley91, Hundley94, Cooley97, Jaime00} is key. 
According to the WKSM scenario, the reduction in spin-orbit coupling (SOC) promoted by replacing Pt with Pd is the main driver
of the destabilization of the Kondo gap in Ce$_{3}$Bi$_{4}$Pd$_{3}$~\cite{Dzsaber17, Lai18}. Dzaber \textit{et al.} found that the isovalent substitution series
Ce$_{3}$Bi$_{4}$(Pt$_{1-x}$Pd$_{x})_{3}$ is also virtually isovolume, which lead to the conclusion that a 
reduction in Kondo coupling, $J_{K}$, by chemical pressure is not the dominant factor in the series~\cite{Dzsaber17}.
In the small-gap insulating scenario, however, Ce$_{3}$Bi$_{4}$Pd$_{3}$ is a reduced $J_{K}$ version of 
Ce$_{3}$Bi$_{4}$Pt$_{3}$ even in the absence of chemical pressure effects~\cite{Kushwaha19}. As illustrated by band structure calculations, an explanation for the reduced $d-f$ hybridization strength along the isovolume series is the reduction in radial extent of the ligand atoms going from $5d$ electrons in Pt to $4d$ electrons in Pd~\cite{Tomczak20}.

 In this Letter, we employ hydrostatic pressure to unravel the narrow-gap insulating ground state of Ce$_{3}$Bi$_{4}$Pd$_{3}$. Here, pressure is an ideal tuning parameter because of its symmetry-preserving nature and ability to increase hybridization while preserving spin-orbit coupling with the advantage of not introducing disorder. Our electrical resistivity measurements show that Ce$_{3}$Bi$_{4}$Pd$_{3}$ behaves as a Kondo insulator under pressure: it becomes progressively more insulating as the hybridization increases, and its insulating gap increases quadratically, similar to the behavior in the series Ce$_{3}$Bi$_{4}$(Pt$_{1-x}$Pd$_{x}$)$_{3}$.

Figure 1 shows the temperature-dependent electrical resistivity, $\rho(T)$, of Ce$_{3}$Bi$_{4}$Pd$_{3}$ at different applied pressures. At low pressures, $\rho(T)$ resembles ambient-pressure data previously obtained in samples polished to remove surface contamination~\cite{Kushwaha19}. At $p = 0.12$~GPa, $\rho(T)$ increases upon decreasing temperature and exhibits an activated behavior at low temperatures. A fit of the data to an Arrhenius formula, $\rho = \rho_{0}\mathrm{exp}(-\Delta/k_{\rm B}T)$, yields a small activated gap of $5~K$, as shown in the inset of Fig.~1. 

\begin{figure}[!ht]
  \begin{center}
  \includegraphics[width=1\columnwidth]{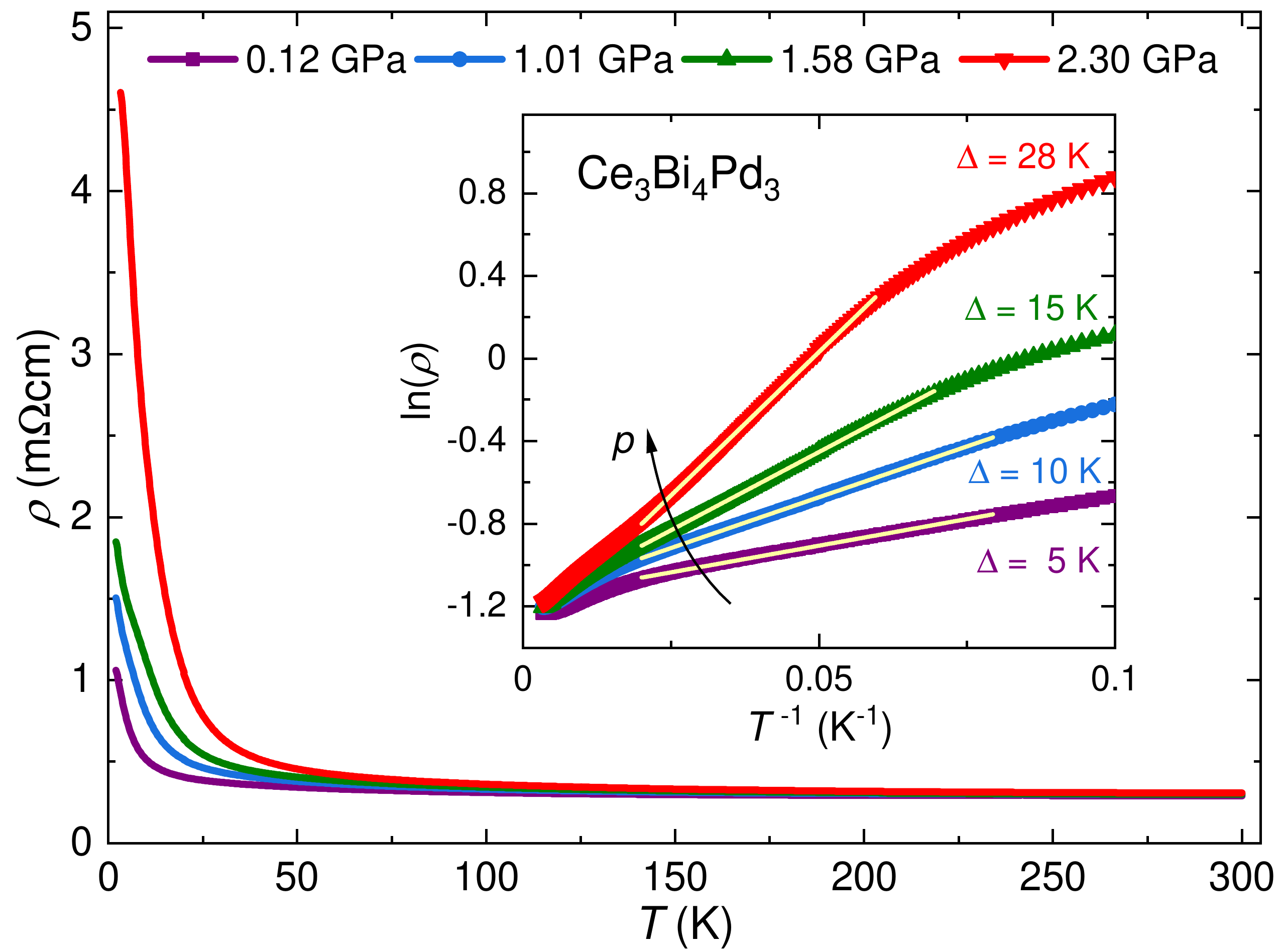}
  \end{center}
  \vspace{-0.5cm}
  \caption{Electrical resistivity of Ce$_{3}$Bi$_{4}$Pd$_{3}$ as a function of temperature at different applied pressures. Inset shows the corresponding Arrhenius plot, ln($\rho$) $vs.$ $1/T$, along with linear fits (yellow lines).}
  \label{fig:Fig1}
  \end{figure}

We note that the magnitude of the extracted gap has to be considered with caution because
the temperature range of the fit is limited to between 50~K and 12~K, which is higher than the gap magnitude. Such a limited temperature range, however, is also observed in the prototypical Kondo insulator Ce$_{3}$Bi$_{4}$Pt$_{3}$~\cite{Hundley90, Hundley91, Hundley94}. The hybridization-driven gap effectively closes at high temperatures ($T>2\Delta/k_{\rm B}$), and the system behaves 
similar to a Kondo metal. The extraction of an activated gap in Ce$_{3}$Bi$_{4}$Pt$_{3}$ is further complicated by its temperature dependence,
which is expected within a mean-field approximation to an Anderson lattice model~\cite{Rise92}. Furthermore, narrow gaps are typically susceptible to even small amounts of defects and impurities at low temperatures, and electrical resistivity measurements are particularly affected by in-gap states~\cite{Sen20}. 
Although early reports attributed the low-temperature resistivity saturation to 
parallel conduction by extrinsic impurity states and variable-range hopping, recent many-body calculations
highlight the role of disorder on intrinsic carriers through finite lifetimes~\cite{pickem20}. In the latter scenario,
a scattering-rate-dependent temperature, $T^*$, marks the crossover from activated behavior at 
modest temperature to resistivity saturation at low temperatures. Nonetheless, estimates of the gap in Kondo insulators from quantities that 
are less dependent on the scattering rate, such as Hall resistivity, magnetic susceptibility, and 
specific heat, yield similar, albeit slightly larger, gap values, e.g., $\approx 20$~K in Ce$_{3}$Bi$_{4}$Pd$_{3}$ at ambient pressure~\cite{Kushwaha19}. 
Therefore, although the magnitude of the extracted gap cannot be taken at face value, 
its pressure dependence is expected to provide qualitative information about the Kondo physics at play.

\begin{figure}[!ht]
\begin{center}
\includegraphics[width=1\columnwidth]{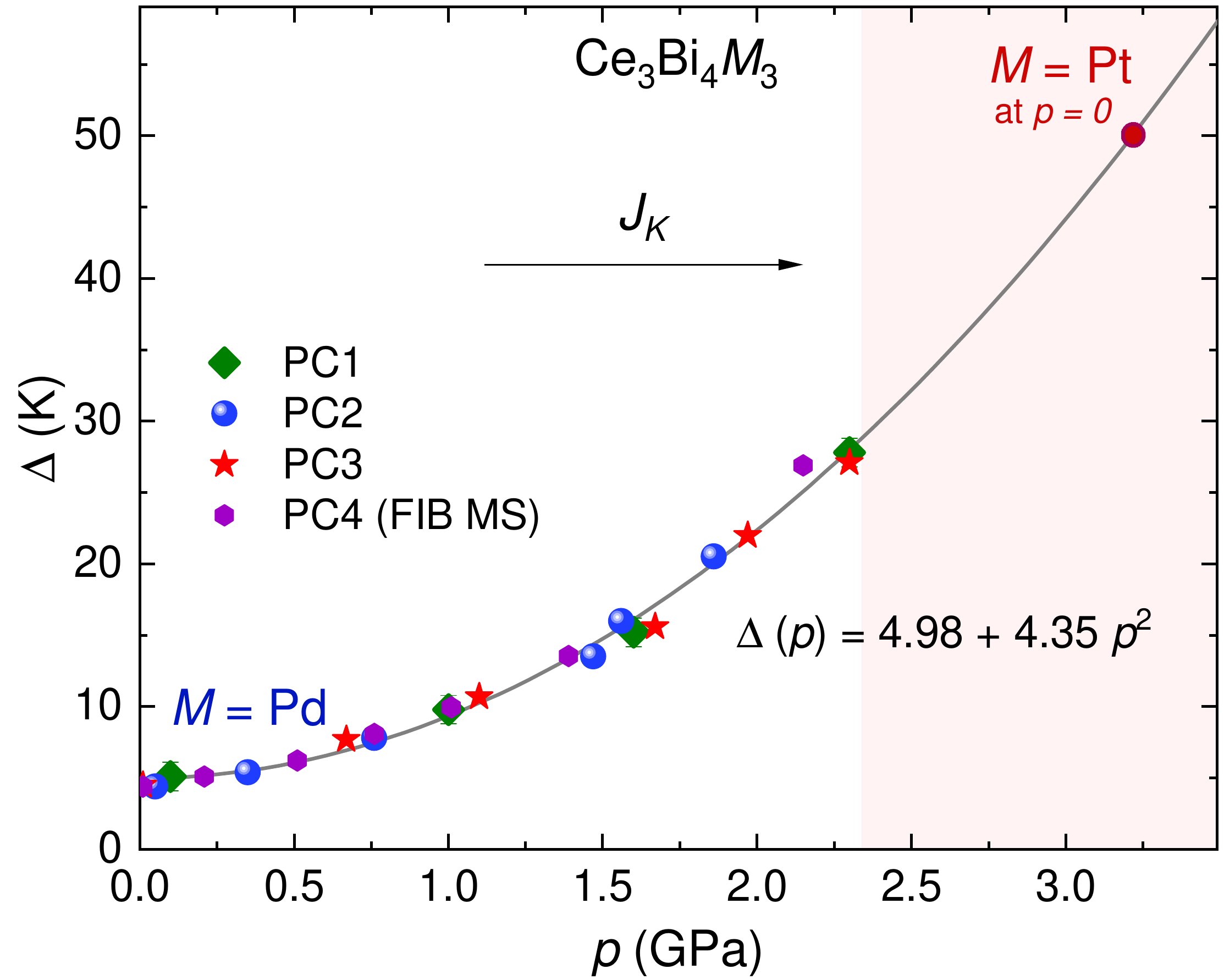}
\end{center}
\vspace{-0.5cm}
\caption{Activated gap $\Delta$ of Ce$_{3}$Bi$_{4}$Pd$_{3}$ as a function of applied pressure; extracted from electrical-resistivity measurements on four different samples. The solid line is a quadratic fit to the data. Pressure increases the Kondo coupling, $J_{\rm K}$, and in turn increases the hybridization gap towards that of Ce$_{3}$Bi$_{4}$Pt$_{3}$.} 
\label{fig:Fig2}
\end{figure}

In fact, the low-temperature resistivity increases monotonically as a function of pressure, which is the first indication that
pressurized Ce$_{3}$Bi$_{4}$Pd$_{3}$ becomes more insulating in a continuous manner. Moreover, at the maximum pressure of our measurements, 
$p = 2.3$~GPa, both the inverse resistance ratio, $\rho_{2\mathrm{K}}/\rho_{300\mathrm{K}}$, and the 
activated gap increase by a factor of 5 compared to the ambient-pressure values. Importantly, Fig.~2 shows that the activated gap increases quadratically with pressure similar to the behavior observed in the substitution series Ce$_{3}$Bi$_{4}$(Pt$_{1-x}$Pd$_{x})_{3}$ with the caveat that no gap was reported for $x=0.37$~\cite{Dzsaber17}, which is possibly caused by increased disorder from substitution. The quadratic increase in $\Delta(p)$ is consistent across several samples measured and a fit to the data (solid line in Fig.~2) provides the pressure dependence of the energy gap as $\Delta(p)=4.98+4.35p^2$.
An extrapolation of this trend to higher pressures indicates that an applied pressure of 3.2~GPa 
would be required to reach the activated gap of the end member Ce$_{3}$Bi$_{4}$Pt$_{3}$, $\Delta \approx 50$~K~\cite{Hundley90}.
Similarly, the activated gap at $p = 2.3$~GPa would correspond to a
stoichiometry of Ce$_{3}$Bi$_{4}$(Pt$_{0.8}$Pd$_{0.2})_{3}$.

In order to understand the effect of pressure on the carrier density, we turn to Hall resistivity measurements. We note that magneto-transport properties at low temperatures are found to be sensitive to impurity inclusions and are strongly sample dependent, even after careful polishing. Possible impurity phases include elemental Bi and Bi-Pd binaries of varying stoichiometry that undergo superconducting transitions at low temperatures~\cite{Roberts76}. A detailed discussion of the extrinsic contributions to the magneto-transport is provided in the Supplementary Information. In order to minimize the contributions from the impurity phases and to obtain well-defined Hall geometry, a microstructure device was fabricated on a thin sample ($600\times250\times50~\mu$m$^3$) of Ce$_{3}$Bi$_{4}$Pd$_{3}$ using focused ion beam. A scanning electron micrograph of the microstructure device is shown in Fig.~3a. Raw Hall resistance data at $p=0.01$~GPa and $T=2$~K is shown in Fig.~3b. The magnetic field dependence of the Hall resistivity, $\rho_{xy}(B)$, measured at $p=0.01$~GPa and different temperatures is presented in Fig.~3c. Here, $\rho_{xy}$ is obtained by antisymmetrizing the resistivity data measured in positive and negative magnetic fields to remove any magnetoresistance contribution from misalignment of the electrical contacts.

\begin{figure}[t]
\begin{center}
\includegraphics[width=1\columnwidth,keepaspectratio]{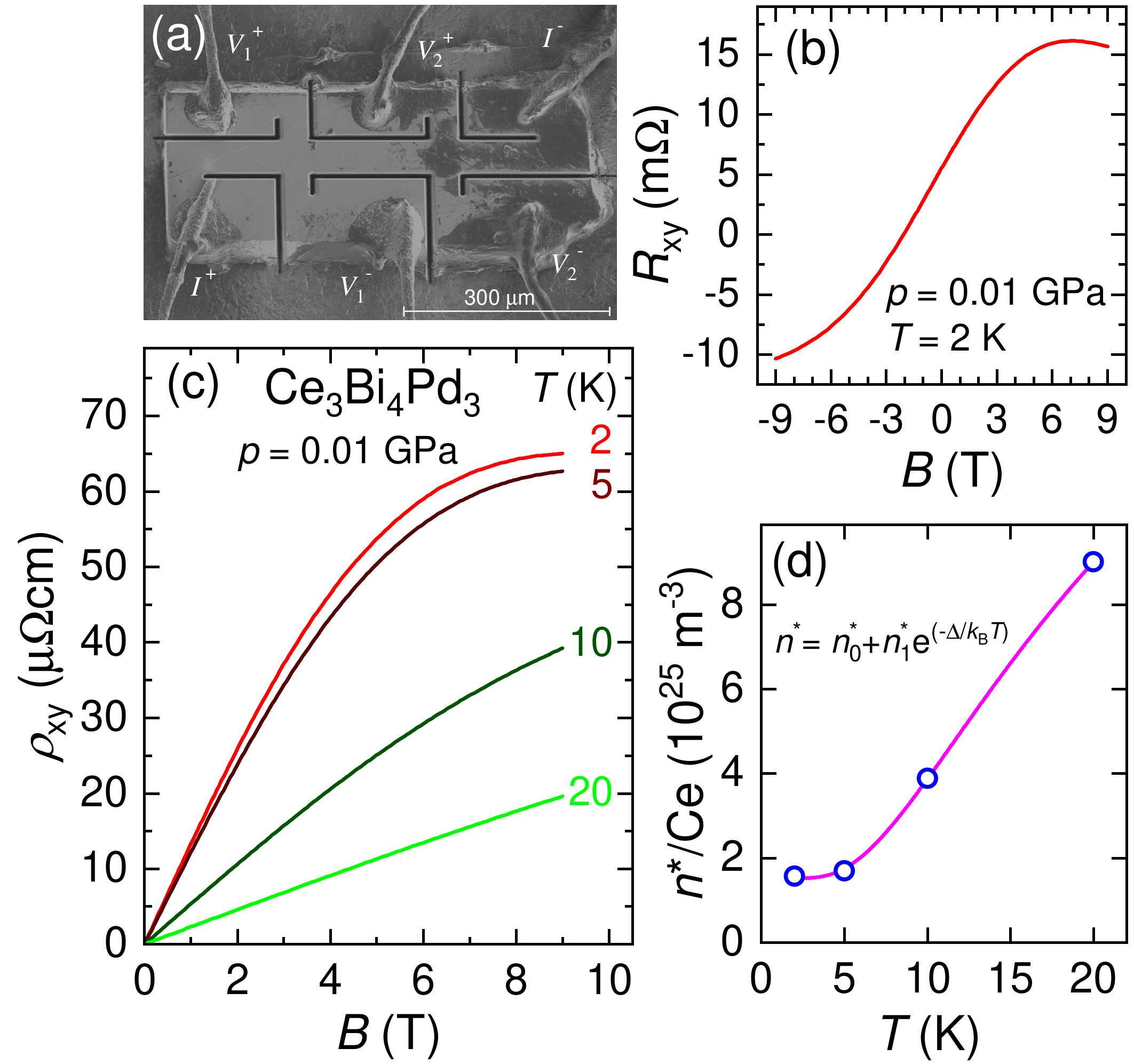}
\end{center}
\vspace{-0.5cm}
\caption{(a) Scanning electron micrograph of the microstructure device used for Hall resistivity measurements. (b) Hall resistance of Ce$_{3}$Bi$_{4}$Pd$_{3}$ at $p=0.01$~GPa and 2~K measured in positive and negative fields. (c) Magnetic field dependence of Hall resistivity $\rho_{xy}(B)$, obtained by antisymmetrizing the as-measured data, of Ce$_{3}$Bi$_{4}$Pd$_{3}$ at $p=0.01$~GPa and different temperatures. (d) The carrier density $n^* = B/e\rho_{xy}$ as a function of temperature (blue circles). The solid line is a fit to the data as described in the main text.}
\label{fig:Fig3}
\end{figure}

At low temperatures, $\rho_{xy}$ increases linearly with magnetic field followed by a downward curvature at higher fields. This is consistent with the results from previous high-field studies which revealed the closing of the Kondo hybridization gap at an applied magnetic field of $B_{c} \approx 11$~T~\cite{Kushwaha19}. The positive $\rho_{xy}(B)$ at low fields indicates that electronic conduction is dominated by hole-like carriers. The linear $\rho_{xy}(B)$ at low fields allow for a straightforward estimation of the carrier density from a single-band model, $n^* = 1/e R_{H}$, where the Hall constant is $R_{H} = \rho_{xy}/B$. The resulting carrier density, calculated from the slope of $\rho_{xy}(B)$ below 2~T, is plotted as a function of temperature in Fig.~3d. Because $n^*$ increases with temperature in a thermally activated manner, it is possible to determine the associated energy gap. A fit to the data (solid line) using $n^* =n_0^*+n_1^*{\rm exp}(-\Delta/k_{\rm B}T)$ yields $n_0^*=1.5\times10^{25}$~m$^{-3}$, $n_1^*=24\times10^{25}$~m$^{-3}$, and $\Delta=23\pm1$~K. The obtained values of carrier density and hybridization gap are in excellent agreement with previous reports~\cite{Kushwaha19, Dzsaber21}.

The evolution of $\rho_{xy}$ vs. $B$, measured at $T=10$~K, with applied pressure is presented in Fig.~4a. It is immediately evident that the slope of $\rho_{xy}(B)$ increases with increasing pressure, suggesting a decrease in carrier density. The carrier density, estimated from the slope of $\rho_{xy}(B)$ below 2~T, is plotted in the inset of Fig.~4a (circles). Such a decrease in $n^*$ with increasing pressure can be understood as the reduction in thermally accessible carriers due to the enhanced energy gap under applied pressure. To further verify this scenario, we calculate $n^*$ as a function of pressure at constant temperature using the pressure dependence of the energy gap $\Delta(p)$ extracted from $\rho(T)$ (see Fig.~2). Here we used the equation for activated behavior $n^* =n_0^*+n_1^*{\rm exp}[-\Delta(p)/k_{\rm B}T]$ along with the pressure dependence of the energy gap $\Delta(p)=23+4.35p^2$ at 10~K. The calculated $n^*(p)$ (solid line in the inset of Fig.~4a) is in excellent agreement with the experimental values obtained from $\rho_{xy}(B)$ data to $p=1$~GPa. We note that determination of the pressure dependence of $n^*$ at lower temperatures is hindered by the effect of impurity phases. In particular, impurity contributions become increasingly prominent at higher temperatures as pressure increases. Above 1 GPa, it is not possible to estimate the intrinsic carrier density even at $T=10$~K (See Supplementary Information). Nonetheless, below 1 GPa the mobility of the carriers can be obtained $via$ the ratio $R_{H}$/$\rho_{xx}$ evaluated at zero magnetic field. The hole mobility, estimated at $T=10$~K, decreases from 104.4~cm$^2$/Vs at $p=0.01$~GPa to 75.1~cm$^2$/Vs at $p=1.01$~GPa. This suggests that the decrease in hole mobility also contributes to the large increase in low temperature resistivity of Ce$_{3}$Bi$_{4}$Pd$_{3}$ at higher pressures.

\begin{figure}[t]
  \begin{center}
  \includegraphics[width=1\columnwidth,keepaspectratio]{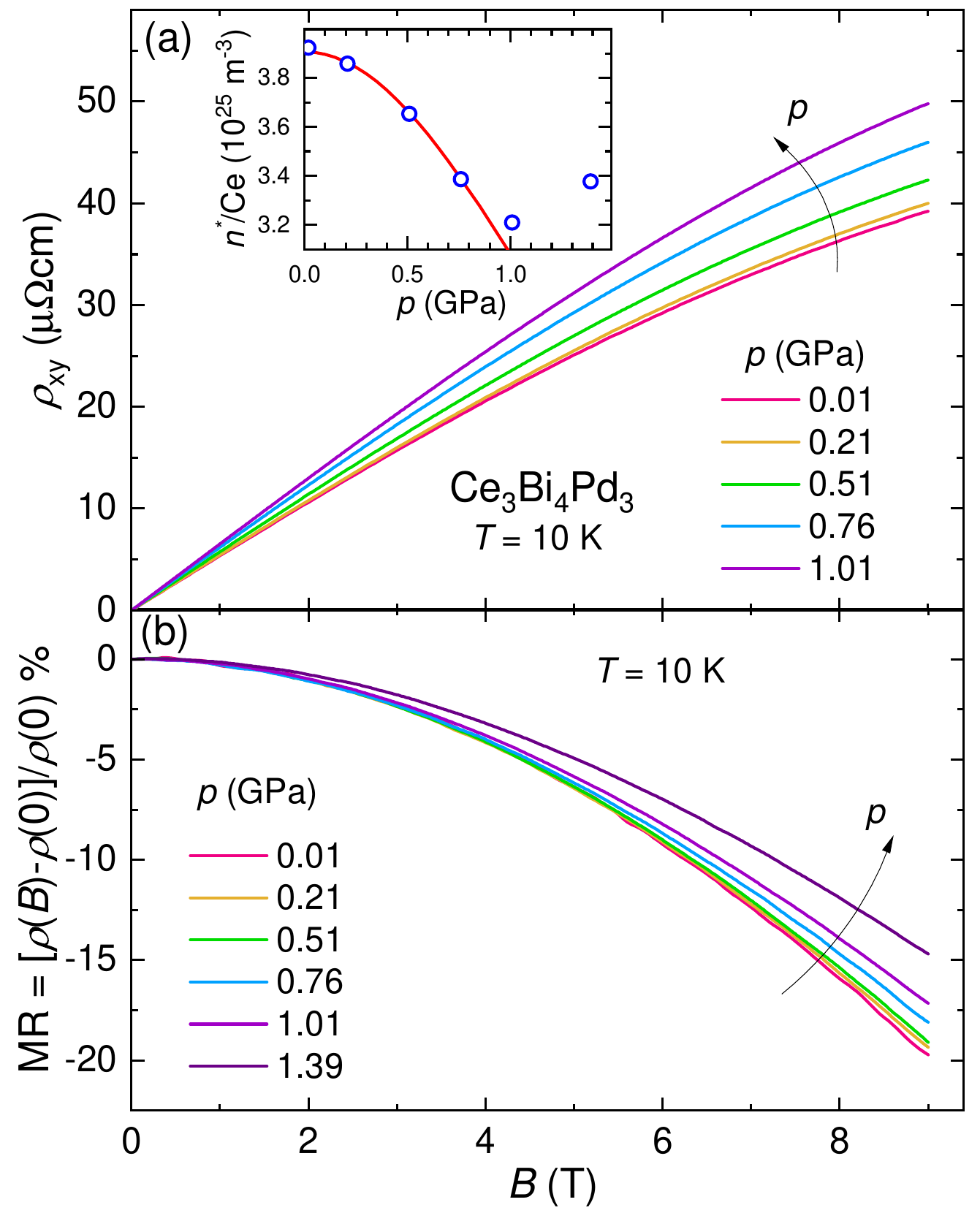}
  \end{center}
  \vspace{-0.5cm}
  \caption{(a) Hall resistivity of Ce$_{3}$Bi$_{4}$Pd$_{3}$ as a function of magnetic fields at $T=10$~K measured at different pressures. The inset shows the pressure dependence of the carrier density $n^*$ (blue circles) where the red line is the expected curve inferred from the pressure dependence of the activated gap $\Delta(p)$ (see main text). (b) Magnetic field dependence of Magnetoresistance MR$(B)=[\rho(B)-\rho(0)]/\rho(0)$ of Ce$_{3}$Bi$_{4}$Pd$_{3}$ measured at $T=10$~K under different applied pressures.}
  \label{fig:Fig4}
\end{figure}

The increase in energy gap with pressure is also evidenced in magnetoresistance. Figure~4b shows the magnetoresistance of Ce$_{3}$Bi$_{4}$Pd$_{3}$, ${\rm MR} =[\rho(B)-\rho(0)]/\rho(0)$,
measured at $T=10$~K. At low pressures, MR is negative and reaches about $-20\%$ at 9~T, in agreement with recent high-field measurements at ambient pressure~\cite{Kushwaha19}. With increasing pressure, the MR$(B)$ curves shift continuously upward, reaching about $-15\%$ at 9~T for $p=1.39$~GPa. Our results are in line with the scenario wherein magnetic fields weaken Kondo coupling and therefore suppress the Kondo hybridization gap. At a critical field of $B_{c} \approx 11$~T, the gap of Ce$_{3}$Bi$_{4}$Pd$_{3}$ was argued to collapse to zero temperature in a way consistent with a quantum critical point. For fields higher than $B_{c}$, a Fermi-liquid regime is subsequently observed in $\rho(T)$~\cite{Kushwaha19}. Because the insulator-to-metal transition is directly related to the Kondo scale, $B_{c}$ is naturally expected to increase as the activated gap increases in Ce$_{3}$Bi$_{4}$Pd$_{3}$ under pressure. Therefore, at higher pressures, larger magnetic fields would be required to reduce the energy gap, resulting in relatively smaller negative MR with increasing pressure.
 
In light of our results under hydrostatic pressure, we now turn to the evaluation of the possible ground states of Ce$_{3}$Bi$_{4}$Pd$_{3}$.
As mentioned earlier, in the WKSM scenario, the reduction in SOC resulting from the substitution of Pt with Pd is argued to be the key tuning parameter that drives the Kondo insulating state in Ce$_{3}$Bi$_{4}$Pt$_{3}$ towards the semimetallic state in Ce$_{3}$Bi$_{4}$Pd$_{3}$~\cite{Dzsaber17}. However, recent band structure calculations suggest that the decrease in SOC arising from the nuclear charge difference between Pt and Pd is insufficient to account for the change in the hybridization gap of the two compounds~\cite{Tomczak20}. Instead the difference in the radial extent of the Pt-$5d$ and Pd-$4d$ orbitals is proposed to significantly affect the hybridization between 4$f$ and conduction electrons and thereby the Kondo coupling, despite the substitution being isovolume. These findings are in line with the results from our hydrostatic pressure study and show that the Kondo coupling is the key parameter governing the ground states in Ce$_{3}$Bi$_{4}$(Pt/Pd)$_{3}$. 

The Kondo coupling strength depends on the hybridization interaction $V$ as $J_{\rm K} \propto |V|^{2}/U$, where $U$ is the Coulomb interaction. Our results in Ce$_{3}$Bi$_{4}$Pd$_{3}$ suggest that the hybridization interaction scales linearly with pressure leading to the observed quadratic pressure dependence of the energy gap. In the substitution series  Ce$_{3}$Bi$_{4}$(Pt$_{1-x}$Pd$_{x}$)$_{3}$, the radial extent of the $d$-orbitals plays a similar role to that of applied pressure. Finally, the striking similarities in the evolution of the Kondo insulating gap with magnetic field and applied pressure in Ce$_{3}$Bi$_{4}$Pd$_{3}$~\cite{Kushwaha19} and Ce$_{3}$Bi$_{4}$Pt$_{3}$~\cite{Cooley97, Jaime00} provide further evidence for a narrow-gap Kondo insulating ground state in Ce$_{3}$Bi$_{4}$Pd$_{3}$.    
 
In summary, we have studied Ce$_{3}$Bi$_{4}$Pd$_{3}$ using electrical transport measurements under hydrostatic pressure. At ambient pressure, Ce$_{3}$Bi$_{4}$Pd$_{3}$ is a narrow-gap Kondo insulator, evidenced by activated behavior in electrical resistivity and Hall measurements. Under hydrostatic pressure, Ce$_{3}$Bi$_{4}$Pd$_{3}$ becomes more insulating, a behavior typical of Kondo insulators such as Ce$_{3}$Bi$_{4}$Pt$_{3}$. Moreover, the hybridization gap shows a quadratic increase with pressure akin to the effect of Pt substitution observed in Ce$_{3}$Bi$_{4}$(Pt$_{1-x}$Pd$_{x}$)$_{3}$ series. The pressure dependence of carrier density, obtained from Hall resistivity measurements on a microstructure device of Ce$_{3}$Bi$_{4}$Pd$_{3}$, as well as the magnetoresistance are in excellent agreement with the increase in energy gap under pressure. Our results suggest that the Kondo coupling prevails as the primary tuning parameter in these materials and Ce$_{3}$Bi$_{4}$Pd$_{3}$ is a narrow-gap version of the prototypical Kondo insulator Ce$_{3}$Bi$_{4}$Pt$_{3}$.  

\begin{acknowledgments}
We acknowledge constructive discussions with Mun K. Chan, Jian-Xin Zhu, and Jan Tomczak. The conception of this work as well the crystal synthesis were supported by the U.S. Department of Energy, Office of Basic Energy Sciences ``Science of 100 T" program.  M. O. Ajeesh acknowledges funding from the Laboratory Directed Research \& Development Program. 
Scanning electron microscope and focused ion beam measurements were supported by the Center for Integrated Nanotechnologies, an Office of Science User Facility operated for the U.S. Department of Energy Office of Science.
Measurements under pressure were supported by the U.S. Department of Energy, Office of Basic Energy Sciences ``Quantum Fluctuations in Narrow Band Systems" program.

\end{acknowledgments}

\subsection{Methods}
Single crystals of Ce$_{3}$Bi$_{4}$Pd$_{3}$ were grown by bismuth flux~\cite{Kushwaha19} and combined bismuth-lead flux technique. In the later, reagents in the starting composition Ce:Pd:Bi:Pb = 1:1:4.25:7 were taken in an alumina crucible placed in an evacuated quartz tube and heated to 800$^0$C in 5 hours and kept there for 24 hours. The solution was slowly cooled down to 650$^0$C in 40 hours and the excess flux was removed by centrifuging. The crystallographic structure was verified at room temperature by a Bruker D8 Venture single-crystal diffractometer equipped with Mo radiation.
Electrical-resistivity measurements were performed in a standard 4-terminal method in which electrical contacts to the samples were made using 12.5 $\mu$m platimum wires and silver paste. The Hall resistivity was measured on a micro-structure device fabricated using Focused Ion Beam (FIB) technique with a SEM SCIOS instrument. A single crystal sample was polished to smaller size ($600\times200\times20$ $\mu$m$^3$), fixed to a sapphire platform ($2\times2$ mm$^2$), and electrical contacts were made using silver paste and 12.5 $\mu$m Pt wires. Micro-patterns were directly milled into the above mentioned sample using Ga ion beam (ion current 65 nA) to fabricate the microstructure device. Electrical-transport measurements under hydrostatic pressure were carried out using a double-layered piston-cylinder-type pressure cell with Daphne 7373 oil as the pressure-transmitting medium. The pressure inside the sample space was determined at low temperatures by the shift of the superconducting transition temperature of a piece of lead (Pb). Electrical resistivity was measured using a AC resistance bridge (Model 372, Lake Shore) at a measuring frequency of 13.7 Hz together with a Physical Property Measurement System (Quantum Design).

\bibliography{Ce3Bi4Pd3}

\end{document}

\title{Supplementary Information for ``Ground state of Ce$_{3}$Bi$_{4}$Pd$_{3}$ unraveled by hydrostatic pressure''}

\author{M. O. Ajeesh$^{1}$, S. M. Thomas$^{1}$, S. K. Kushwaha$^{2}$, E. D. Bauer$^{1}$, F. Ronning$^{1}$, J. D. Thompson$^{1}$, N. Harrison$^{2}$, and P. F. S. Rosa$^{1}$}
\affiliation{
$^{1}$  Los Alamos National Laboratory, Los Alamos, New Mexico 87545, U.S.A.\\
$^{2}$ National High Magnetic Field Laboratory, Los Alamos, New Mexico 87545, U.S.A.}
\date{\today}

\maketitle

\subsection{Effect of impurity inclusions}

 The low-temperature transport properties of Ce$_{3}$Bi$_{4}$Pd$_{3}$ are extremely sample dependent. For instance, previously reported low-temperature resistivity values of Ce$_{3}$Bi$_{4}$Pd$_{3}$ vary by almost an order of magnitude~\cite{Dzsaber17, Kushwaha19, Dzsaber21}. Further, in one report, certain samples showed a drop in resistivity around $T=3$~K due to surface contamination that could be removed by careful polishing (see Supplementary Figure 3 of Ref.~\cite{Kushwaha19}). Possible impurity phases include elemental bismuth as well as Bi-Pd binary compounds of varying stoichiometry. Several of these binary compounds are superconducting at low-temperatures: tetragonal Bi$_2$Pd ($T_{\rm c}=4.25$~K), monoclinic Bi$_2$Pd ($T_{\rm c}=1.73$~K), BiPd ($T_{\rm c}=3.74$~K), Bi$_{0.33}$Pd$_{0.67}$ ($T_{\rm c}=4$~K), and Bi$_{0.4}$Pd$_{0.6}$ ($T_{\rm c}=3.7-4$~K)~\cite{roberts76}. In most samples investigated here, low-temperature resistivity showed clear effects of superconducting impurities which could not be removed by polishing. This suggests that the impurity phases could also reside within the bulk of the samples as inclusions, which is a known disadvantage of crystal growth through the flux method.
 
 \begin{figure}[t!]
  \begin{center}
    %\vspace{-0.2cm}
  %\hspace{-0.35cm}
  \includegraphics[width=0.8\columnwidth]{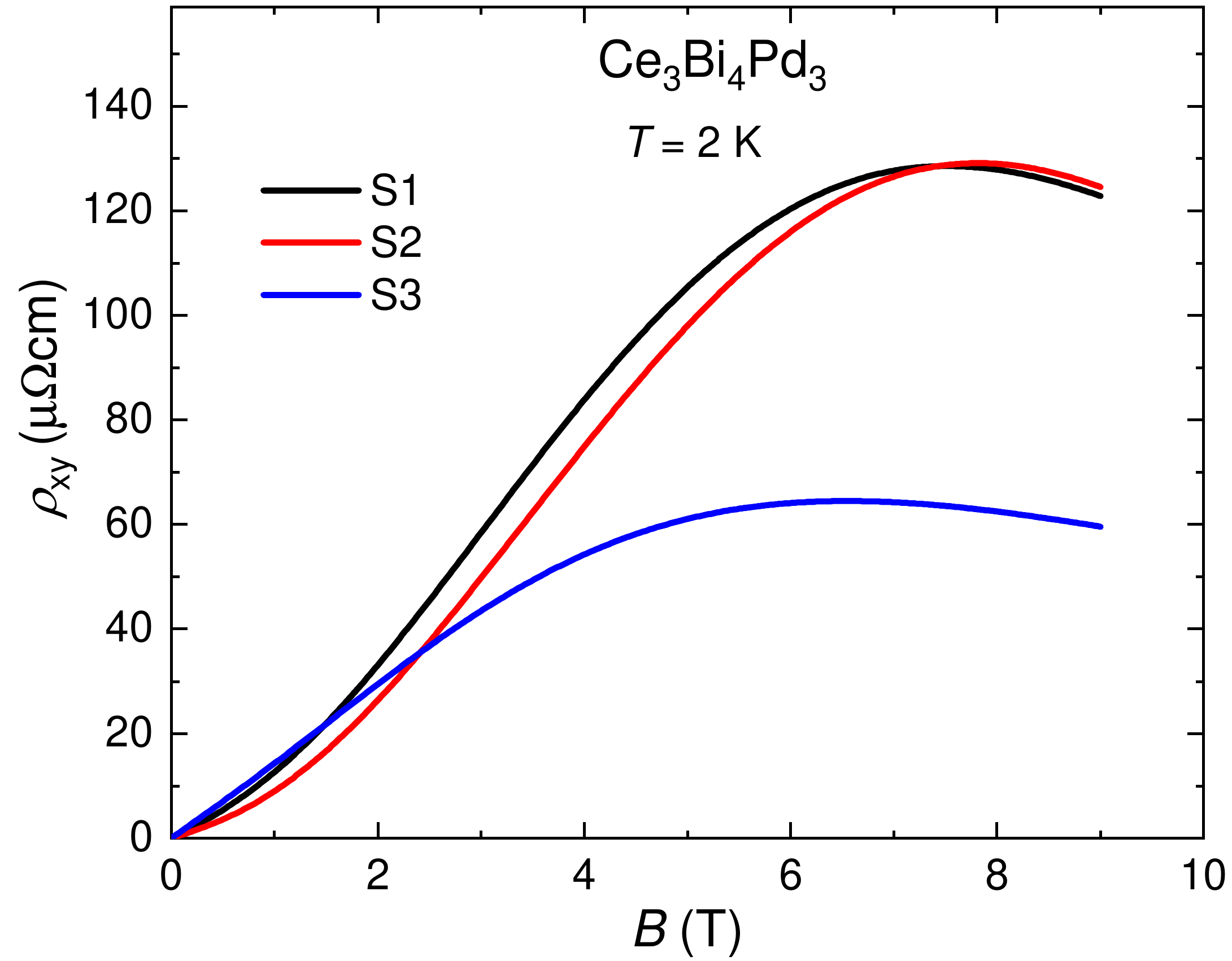}
  \vspace{-0.5cm}
  \end{center}
  \caption{Hall resistivity  as a function of magnetic field measured at $T=2$~K on three different polished samples of Ce$_{3}$Bi$_{4}$Pd$_{3}$.}
  \label{fig:Fig1}
  %\vspace{-0.2cm}
  \end{figure}

Impurity inclusions cause particularly pronounced effects in the Hall resistivity. Contributions from the impurity inclusions appear as non-linear behavior in the Hall resistivity $\rho_{\rm xy}(B)$. In Fig.~S1, $\rho_{\rm xy}(B)$ measured at $T=2$~K for three different polished samples of Ce$_{3}$Bi$_{4}$Pd$_{3}$ is presented. Samples S1 and S2 show non-linear $\rho_{\rm xy}(B)$ in the low-field region, whereas sample S3 shows linear $\rho_{\rm xy}(B)$ and the Hall resistivity values are consistent with reported values~\cite{Kushwaha19}. Therefore, we use the linear behavior in $\rho_{\rm xy}(B)$ to select samples with less pronounced impurity inclusions.

\begin{figure}[t!]
  \begin{center}
    %\vspace{-0.2cm}
  %\hspace{-0.35cm}
  \includegraphics[width=1\columnwidth]{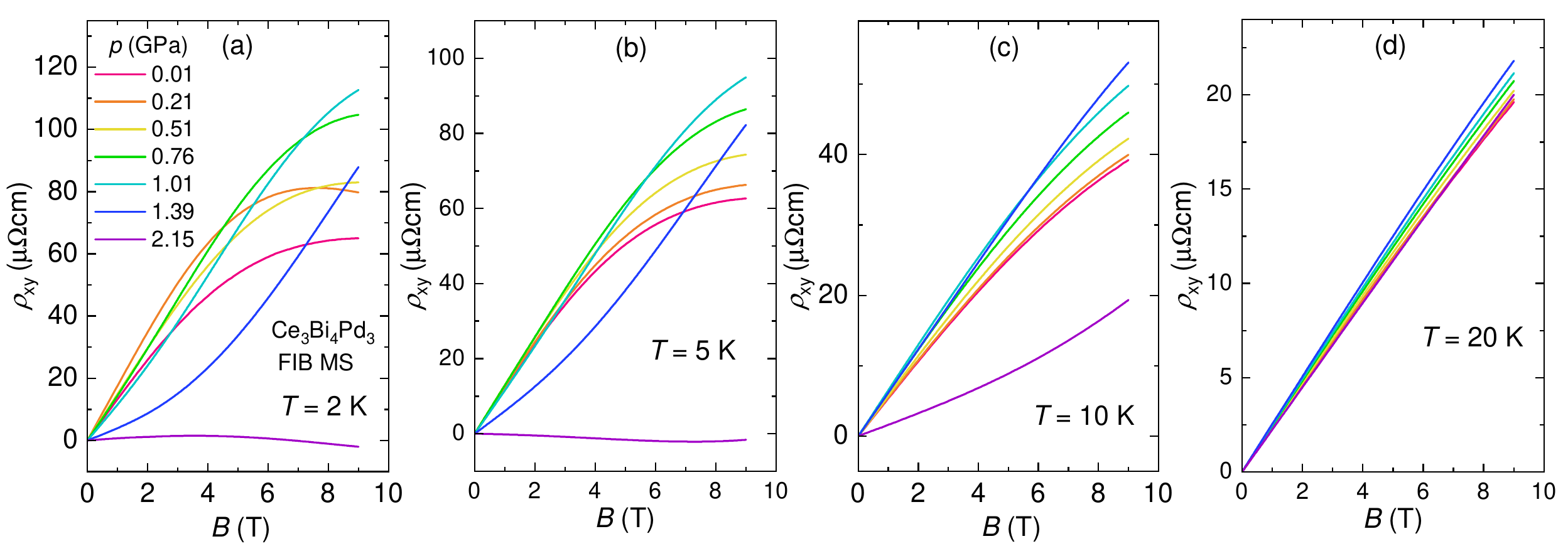}
  \vspace{-0.5cm}
  \end{center}
  \caption{Hall resistivity of Ce$_{3}$Bi$_{4}$Pd$_{3}$ as a function of magnetic field measured at different applied pressures at temperature of (a) 2~K (b) 5~K (c) 10~K (d) 20~K; measured on microstructure device.}
  \label{fig:Fig1}
  %\vspace{-0.2cm}
  \end{figure}

The non-linear behavior in Hall resistivity becomes increasingly pronounced under the application of pressure. Surprisingly, this is also the case for a microstructure device fabricated using a small polished crystal. Hall resistivity as a function of magnetic field measured under several pressures for different temperatures is plotted in Fig.~S2. At $T=2$~K, even though the low pressure $\rho_{\rm xy}(B)$ is linear, non-linearity begins to appear for $p>0.5$~GPa. Similar behavior is also observed at higher temperatures, where the non-linear behavior sets in at relatively higher pressures. At $T=10$~K, $\rho_{\rm xy}(B)$ remains linear in the low-field region for pressures to 1~GPa, therefore allowing the determination of the pressure dependence of the intrinsic carrier density of Ce$_{3}$Bi$_{4}$Pd$_{3}$ (see Fig.~4a in the main text).

\begin{figure}[t!]
\begin{center}
\includegraphics[width=1\columnwidth]{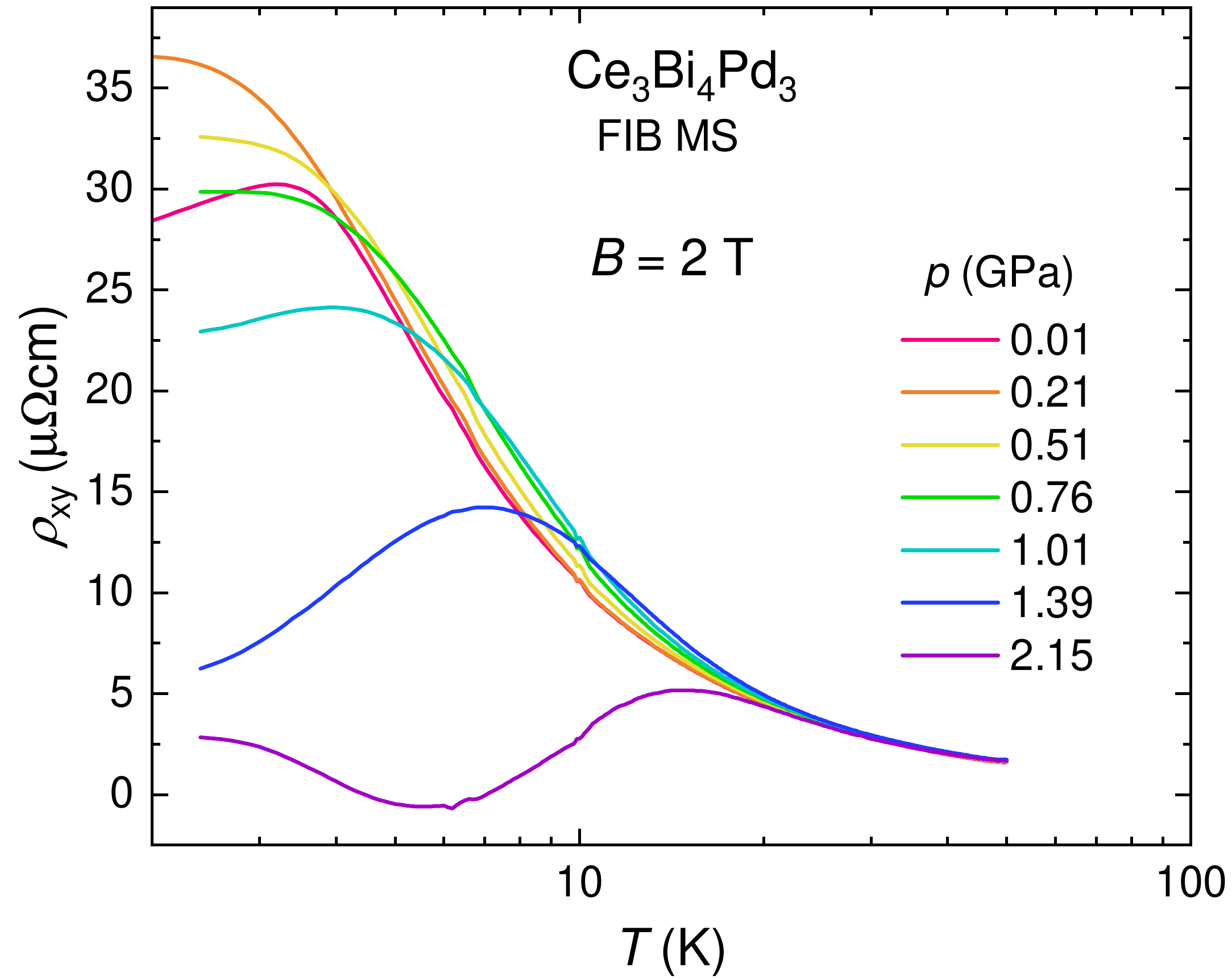}
\end{center}
%\vspace{-0.5cm}
\caption{Temperature dependence of Hall resistivity of Ce$_{3}$Bi$_{4}$Pd$_{3}$ for several applied pressures; Hall resistivity is obtained as $\rho_{\mathrm xy}(T)=[\rho_{\mathrm +2T}(T)-\rho_{\mathrm -2T}(T)]/2$}
\label{fig:Fig2}
\end{figure}

The observation of impurity contributions becoming increasingly prominent at higher temperatures with increasing pressure is better visualized in the temperature-dependent Hall resistivity measured at different pressures. The temperature dependence of $\rho_{\rm xy}$ is obtained by antisymmetrizing the data measured in $B=+2$~T and $B=-2$~T. The feature associated with the impurity contribution to the Hall resistivity shifts to higher temperatures with increasing pressure, as seen in Fig.~S3. At the highest pressure in our study, $p=2.15$~GPa, the effect from impurity contributions can be found up to $T=30$~K.

\begin{figure}[t!]
\begin{center}
\includegraphics[width=0.6\columnwidth,keepaspectratio]{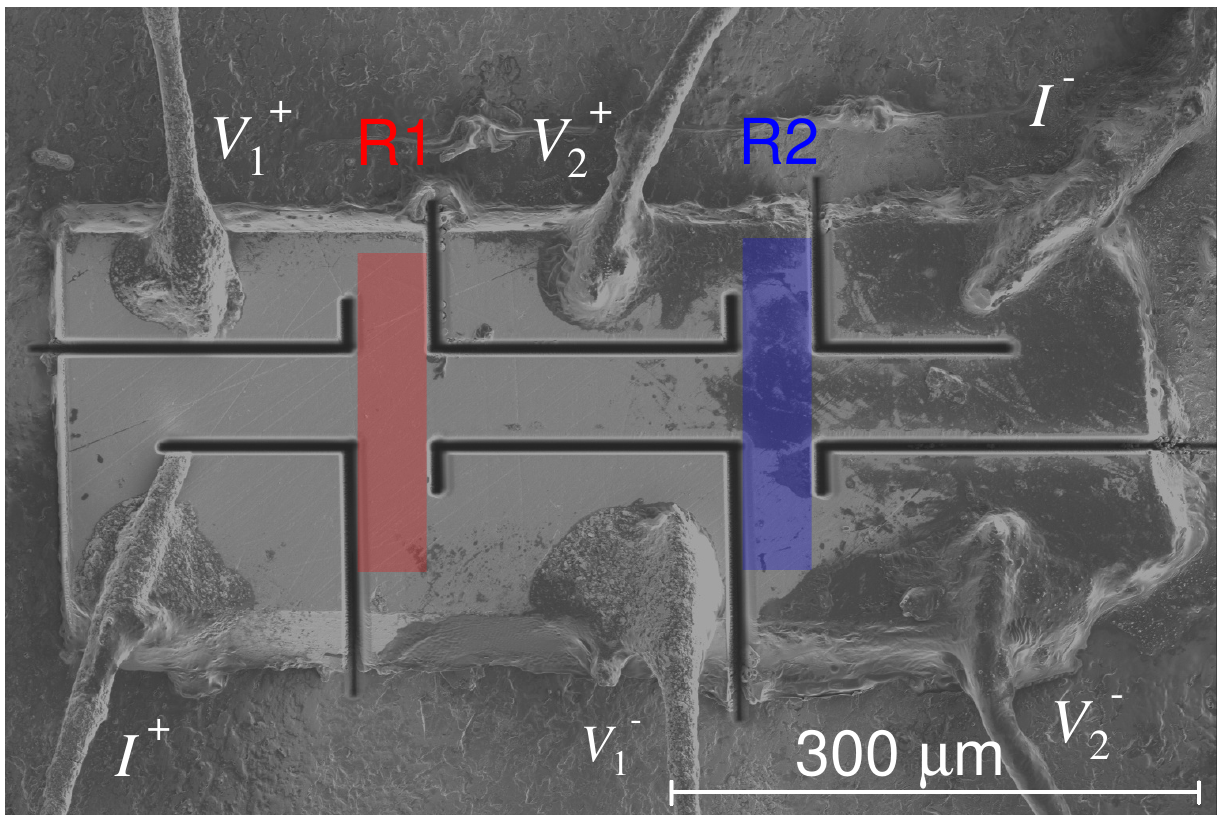}
\end{center}
%\vspace{-0.5cm}
\caption{Scanning electron micrograph of the microstructure device of Ce$_{3}$Bi$_{4}$Pd$_{3}$ used for Hall resistivity measurements. Two separate Hall channels, marked by red (R1) and blue (R2) areas, are measured.}
\label{fig:Fig3}
\end{figure}

\begin{figure}[t!]
  \begin{center}
  \includegraphics[width=1\columnwidth,keepaspectratio]{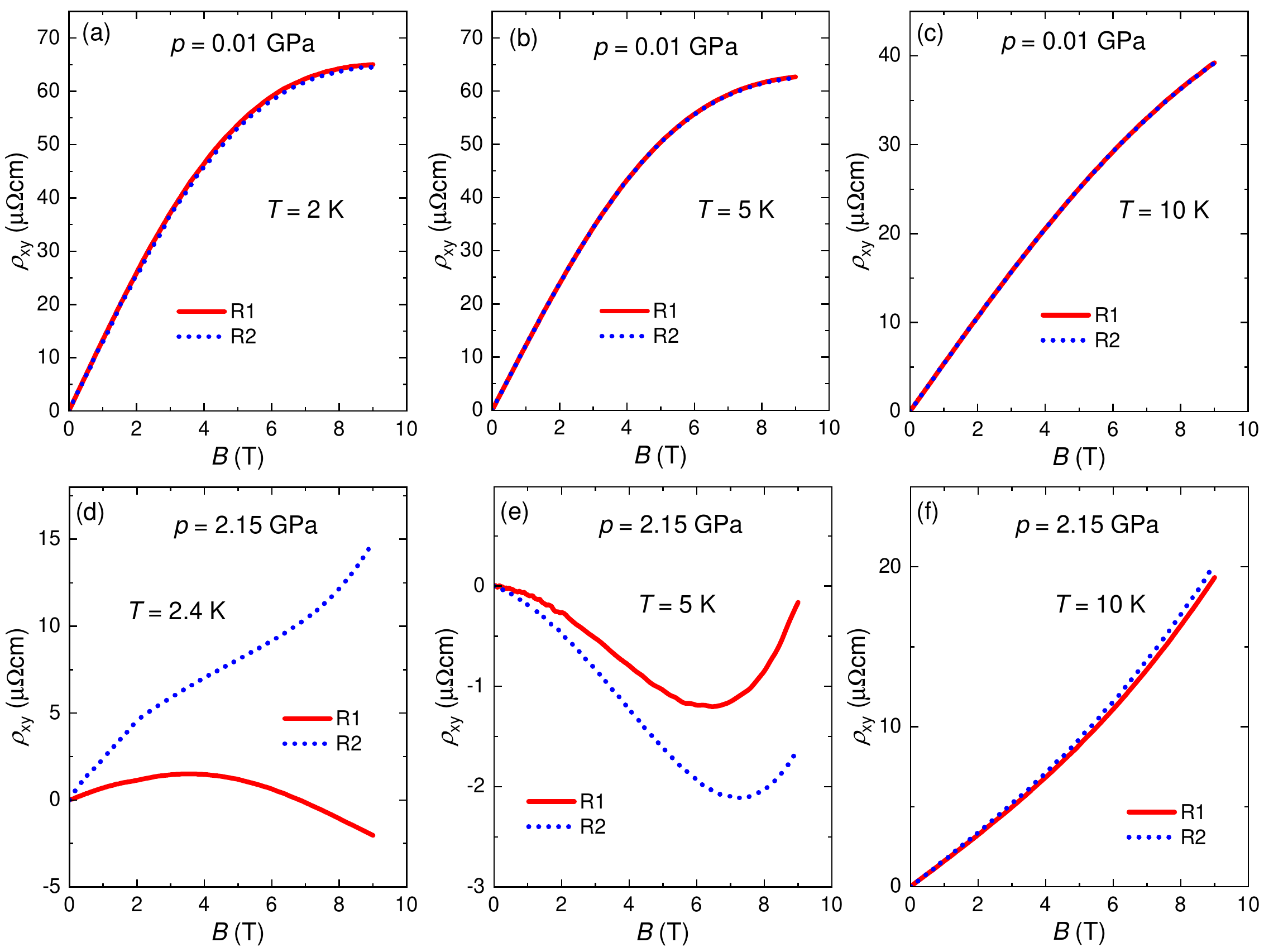}
  \end{center}
  %\vspace{-0.5cm}
  \caption{Comparison of the Hall resistivity measured using the two Hall channels R1 (solid red line) and R2 (dotted blue line) of the microstructure device of Ce$_{3}$Bi$_{4}$Pd$_{3}$. $\rho_{\rm xy}(B)$ measured at different temperatures for $p=0.01$~GPa (top panels) and $p=2.15$~GPa (bottom panels).}
  \label{fig:Fig4}
\end{figure}

In general, non-linear behavior in Hall resistivity can develop under pressure due to changes in the band structure. In contrast, non-linear behavior could also arise due to the largely unknown pressure dependence of the various possible inclusions. To verify whether the observed non-linear Hall resistivity under pressure is intrinsic, we compared the Hall resistivity measured in two different channels in the same sample. To this end, we fabricated a microstructure device with two Hall channels with identical geometries (marked R1 and R2 in Fig.~S4). A comparison of the Hall resistivity measured in the two channels at different temperatures for $p=0.01$~GPa (top panel) and $p=2.15$~GPa (bottom panel) is presented in Fig.~S5. At $p=0.01$~GPa, the Hall resistivities of both channels fall on top of each other. Notably, at $p=2.15$~GPa, the Hall resistivities of the two channels show remarkable differences, especially at low temperatures. This result provides strong evidence that the non-linear contributions in the Hall resistivity is primarily extrinsic in origin.  

Even though we find conclusive evidence for extrinsic contributions from impurity inclusions, even a qualitative understanding of such contributions in magneto-transport properties is difficult. First, these inclusions may be present in the form of irregular filaments, which in turn would give rise to current inhomogeneities. Second, as mentioned above, the inclusions could range from elemental bismuth to Bi-Pd binary compounds of varying stoichiometry. Such binaries have a range of superconducting transition temperatures that could also evolve differently with application of pressure. Third, the normal state of such binaries often have semimetallic behavior with strong magnetoresistance properties. In the case of elemental Bi, strongly anisotropic magneto-transport behavior is observed along with a change from semimetallic to semiconducting behaviour at low-temperature with increasing pressure~\cite{Li17}. A semimetal to insulator-like behavior change is also observed under magnetic field applied along the trigonal axis of Bi single crystal~\cite{Du05}. Effects can be further complicated as the inclusions can have geometries such as thin-layer or filament shape which could lead to different properties compared to bulk samples~\cite{Chu92, Chu88, Lu96}. 
%Moreover, the distribution as well as the orientation of these inclusions within the sample could result in complicated response in magneto-transport measurements.
Based on these results, utmost care must be taken in interpreting results from magneto-transport measurements on samples with possible inclusions of bismuth and associated binary compounds.

%\begin{thebibliography}{99}

\bibliography{Ce3Bi4Pd3}

%\end{thebibliography}